# THz emission from Co/Pt bilayers with varied roughness, crystal structure, and interface intermixing

G. Li[1*†], R. Medapalli[3*], R.V. Mikhaylovskiy[2], F.E. Spada[3], Th. Rasing[1], E. E. Fullerton[3], and A.V. Kimel[1,4]

[1] Radboud University, Institute for Molecules and Materials, Heyendaalseweg 135, Nijmegen, The Netherlands.

[2] Department of Physics, Lancaster University, Bailrigg, Lancaster LA1 4YW, United Kingdom.

[3] Center for Memory and Recording Research, University of California, San Diego, La Jolla, California 92093-0401, USA.

[4] Moscow Technological University, MIREA, Vernadsky Ave. 78, Moscow 119454, Russia.

**Ultrafast demagnetization of Co/Pt heterostructures induced by a femtosecond 800 nm laser pulse launches a spin-current from Co to Pt and subsequent conversion of the spin-current to a charge current in the Pt-layer due to the inverse spin-Hall effect. At the same time, due to the spin-dependent photogalvanic effect, a circularly polarized femtosecond laser pulse also generates a photocurrent at the Co/Pt interface. Both ultrashort photocurrent pulses are effectively detected in a contactless way by measuring the THz radiation they emit. Here we aim to understand how the properties of the Co/Pt interface affect the photocurrents in the bilayers. By varying the interfacial roughness, crystal structure and interfacial intermixing as well as having an explicit focus on the cases when THz emissions from these two photocurrents reveal opposite trends, we identify which interface-properties play a crucial role for the photocurrents. In particular, we show that reducing the roughness, the THz emission due to the spin-dependent photogalvanic effect reduces to zero, while the strength of the THz emission from the photocurrent associated with the inverse spin-Hall effect increases by factor of 2. On the other hand, while intermixing strongly enhances the THz emission from the inverse spin-Hall effect by a factor of 4.2, THz emission related to the spin-dependent photogalvanic effect reveals the opposite trend. These findings indicate that microstructural properties of the Co-Pt interface plays a decisive role in the generation of photocurrents.**

## (1) Introduction

For the last two decades, optical manipulation and excitation of magnetization dynamics at the picosecond and sub-picosecond timescales has been an active research field in magnetism [1-3], triggered by the seminal observation of ultrafast demagnetization of a thin Ni-film by a sub-100-fs laser pulse [4]. This demagnetization occurred much faster than any elementary interactions involving spins known at the time. Together with an intense search for the mechanisms of the ultrafast demagnetization, these first experiments also launched debates about the role of artifacts in ultrafast time-resolved measurements and the validity of the conclusions of these initial experiments [5-6]. These discussions motivated the development of new probes for ultrafast magnetization dynamics and resulted, in particular, in an elegant proposal to employ THz time-domain emission spectroscopy, which relies on the fact that any magnetic dipole change must be accompanied by an emission of electromagnetic radiation [7-9]. However, the application of THz time-domain emission spectroscopy for the study of ultrafast magnetization dynamics in magnetic multilayers led to rather unexpected results – the THz electric field emitted as the result of the laser induced demagnetization of Fe/Au and Fe/Ru bilayers was stronger than the THz electric field emitted as the result of the demagnetization of a single Fe-film [10]. It was shown that ultrafast demagnetization generates a spin-polarized current from Fe to Pt, where, as a result of strong spin-orbit interaction, the spin-polarized current pulse was transformed to a charge current pulse that is a more efficient source of THz emission than the magnetization dynamics itself. This mechanism for spin-to-charge current conversion is known as the inverse spin-Hall effect. The photocurrent and THz emission did not depend on the polarization of the excitation light and the phase of the emitted THz radiation could be changed by 180-degrees by a magnetization reversal. These findings resulted in the development of new THz emitters based on spin-polarized currents [11-14].

* Both authors performed equal amounts of work into the paper
† Contact: Qiao.Li@science.ru.nl

The first experiments on ultrafast laser-induced demagnetization, which did not depend on the polarization of light, also inspired the search for ultrafast polarization-dependent effects of light on magnetism. Although many groups tried, demonstration of an ultrafast polarization-dependent effect of light on magnetic materials remained elusive, raising doubts about the feasibility of this phenomenon [15-17]. The first effect of circularly polarized femtosecond laser pulses on spins in a magnetic medium was discovered in the dielectric canted antiferromagnetic $DyFeO_3$ and later demonstrated in a broad class of materials, including dielectric compensated antiferromagnets, dielectric and metallic ferrimagnets as well as ferromagnetic semiconductors [18-22]. Also in the case of polarization-dependent effects of light on spins, THz time-domain emission spectroscopy was shown to be a powerful tool, demonstrating the vectorial control of spins in antiferromagnetic NiO [23]. In Ref. [24], it was demonstrated that polarization-dependent THz emission from ferromagnetic Co-films can be dramatically enhanced by growing a thin capping layer of Pt on top of the Co. The effect was explained to result from a helicity-dependent femtosecond laser-induced rotation of the magnetization of Co in the plane of the sample which, due to the inverse spin-orbit torque, generates a femtosecond pulse of electric current at the Co/Pt interface. Generation of a photocurrent at the interface can be also seen as a spin-dependent photogalvanic effect, reported earlier for noncentrosymmetric semiconductors in an external magnetic field [25]. This current generates a polarization-dependent emission of THz radiation. This observation opens up appealing opportunities for fundamental studies of THz spintronics. In conventional spintronic devices, the direction of the current is controlled by both voltage and magnetic field. At the same time, Huisman *et al.* [24] showed that the direction of sub-picosecond current pulses can be changed in a contactless manner, without applying a voltage, by simply changing the helicity of the excitation light.

Although it is clear that the polarization-dependent and polarization-independent photocurrents in Co/Pt bilayers should depend on the properties of the interface between the Co and Pt layer, it still remains unclear if and to what extent the roughness, crystal structure, and intermixing at this interface play any role in the process of the photocurrent generation. As mentioned above, both ultrashort photocurrent pulses are effectively detected in a contactless way by measuring the THz radiation emitted. Here, by varying the roughness, crystal structure, and interface intermixing as well as explicitly focusing on the cases when THz emissions from these two photocurrents reveal opposite trends, we identify which interfacial properties play a crucial role for the generation of photocurrents. We first fabricated polycrystalline Co/Pt bilayers by magnetron sputtering where the Co layer was deposited at various pressures to vary the microstructure and corresponding roughness at the Co/Pt interface. The quality of the interface was varied from a smooth to a relatively rough interface, by varying the deposition chamber Ar pressure for depositing the Co layer from 3 mTorr to 40 mTorr. We then fabricated Co/Co$_x$Pt$_{1-x}$/Pt bilayers where the Co and Pt layers are separated by a 1-or 2-nm-thick Co$_x$Pt$_{1-x}$ alloy spacer layer with x being 0.25, 0.50, or 0.75. In these samples we studied the role of intermixing between Co and Pt atoms at the interface. Finally, we fabricated Co/Pt via epitaxial sputtering deposition where the Co was grown in the face centered cubic (FCC) or hexagonal close packing (HCP) structure. In these samples we studied the role of the crystal structure and magnetic anisotropy on the observed laser-induced HI and HD THz emission. We found that the interfacial roughness affects the HI THz emission, but more importantly, plays a crucial role for the generation of the HD THz emission. We also found that the Co$_x$Pt$_{1-x}$ alloy spacer layer, *i.e.* intermixing, has an amplification effect on the HI THz emission, but no effect on the HD THz emission. In addition, we did find a strong dependence of the HI THz emission in HCP Co/Pt with respect to the easy axis. For both the FCC and HCP Co/Pt we found little difference in the THz emission compared to the textured Co/Pt bilayer grown at 3 mTorr.

## (2) Sample Fabrication and Characterization

### (a) Interfacial Roughness Series

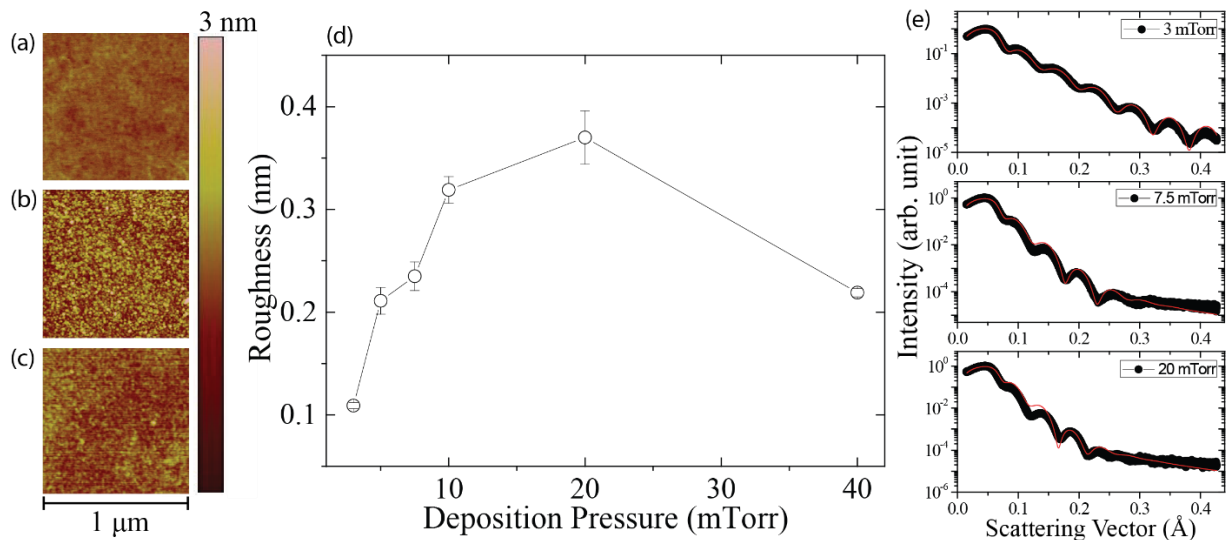

*Figure 1 AFM images of the surface of sputter-deposited pure Co-films grown on glass substrates at an Ar deposition pressure of (a) 3 mTorr, (b) 20 mTorr, and (c) 40 mTorr. (d) AFM root-mean-square roughness of the surface of the Co-films grown at various deposition pressures. (e) The X-ray reflectometry signal of Co/Pt bilayers*

*where the Co layer was grown at deposition pressures 3, 7.5, and 20 mTorr with the intensity plotted in logarithmic scales. The red solid curves are the fits to the experimental data.*

Using magnetron sputtering we fabricated polycrystalline Co/Pt bilayers on soda lime glass with varying interface roughness. For our study we fixed the thickness of the Co and Pt layers to 10 and 3 nm, respectively. Simultaneously, we fabricated Co-films (without Pt layer) for reference studies. To achieve a precise control over the roughness at the Co/Pt interface, for the growth of the Co layer we varied the Ar sputtering pressure from 3-40 mTorr. As the sputtering pressure increases, scattering of the sputtered Co atoms with the Ar atoms in the gas leads to a reduction of the average kinetic energy of the deposited atoms and the motion of Co atoms becomes more random. This results in a low surface mobility of the deposited atoms at high pressures and increased effects of self-shadowing. As the deposition evolves dome shaped grains are formed with well-defined grain boundaries known as zone-1 growth in thin-film deposition [26,27]. For the Pt layer, we kept the Ar-sputter pressure at 3 mTorr to form a dense capping layer. We find that the Co interfacial roughness increases with Ar pressure as expected. Figures 1(a-c) show the atomic force microscopy (AFM) images of the reference pure Co samples without the Pt capping layer. These pure Co-films were grown at deposition pressures 3, 5, 7.5, 10, 20, and 40 mTorr, revealing the change in surface texture from a smooth flat to a high roughness profile. Figure 1(d) shows the evolution of the AFM root-mean-square roughness of the Co surface with Ar sputtering pressure, showing a monotonic increase the roughness with Ar sputter pressure.

The X-ray reflectometry measurements on the Co/Pt bilayers shown in Fig. 1(e) are consistent with the AFM results up to 20 mTorr. The strong suppression in the oscillations at higher angles for samples grown at higher deposition pressures confirms that interfacial roughness indeed increases with increasing deposition pressure. For the samples grown at the deposition pressures 30 and 40 mTorr the interfaces were sufficiently rough that the X-ray reflectivity oscillatory signal was completely suppressed. This is consistent with the expected significantly higher roughness compared to the 3-mTorr Co layer.

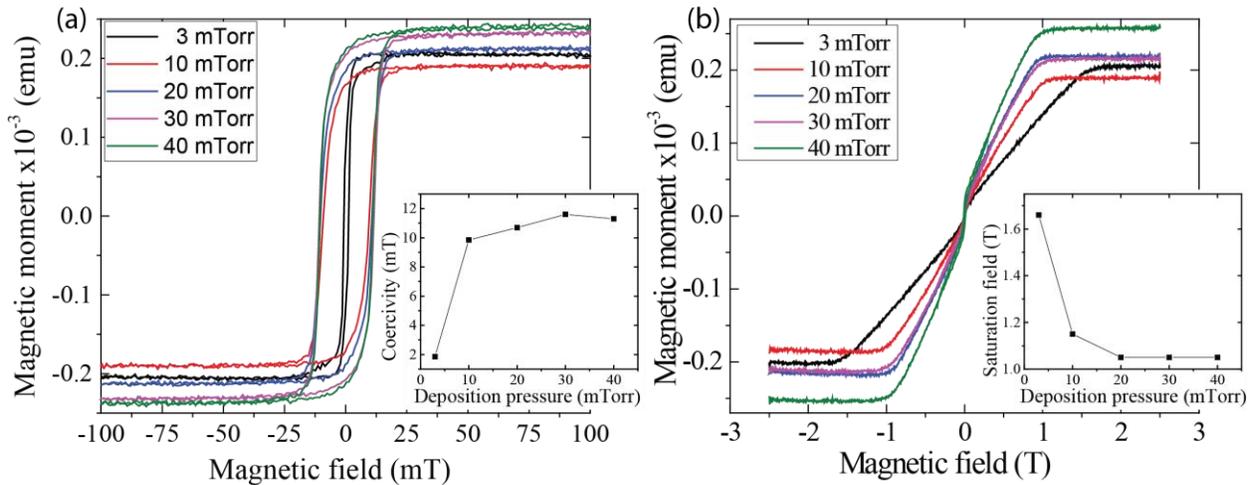

*Figure 2 (a)In-plane magnetic hysteresis loops for Co/Pt bilayers measured in Co/Pt bilayers grown at various deposition pressures of Co with the help of a vibrating sample magnetometer. (inset) The coercive field of each hysteresis loop measured in the in-plane magnetic field. (b) The corresponding out-of-plane magnetic hysteresis*

*loop measurements are shown. (inset) The strength of the applied out-of-plane magnetic field corresponding to the saturation magnetization.*

For the magnetic characterization of the samples with varying roughness we used vibrating sample magnetometry and measured the magnetic hysteresis loop in an external magnetic field (see Fig. 2). The measurements were performed with the external magnetic field oriented either in-plane or out-of-plane of the sample. Figures 2(a, b) summarize the in-plane and out-of-plane magnetometry measurements, respectively, obtained for Co/Pt bilayers grown at 3 mTorr up to 40 mTorr deposition pressures. The in-plane magnetization can be reversed and saturated by applying a rather small magnetic field of 12 mT, while the out-of-plane magnetization requires a much stronger applied magnetic field of more than 1.7 T. The inset of Fig. 2(a) summarizes how the in-plane coercive field depends on the deposition pressure. The inset of Fig. 2(b) shows how the deposition pressure affects the strength of the out-of-plane magnetic field corresponding to the saturation of the magnetization.

The magnetic properties change drastically between samples that were grown at a deposition pressure of 3 mTorr and 10 mTorr, as reflected in Fig. 2. Firstly, for Co/Pt grown at 3 mTorr and 10 mTorr the in-plane magnetic coercive field increases from 1.85 mT to 9.85 mT, respectively. The samples grown at deposition pressures larger than 10 mTorr show smaller changes in the coercive fields [28]. This is expected as the well-defined grain boundaries in high-pressure growth can effectively pin the domain walls during reversal. Secondly, the out-of-plane magnetic saturation for Co/Pt grown at 3 mTorr and 10 mTorr drops from 1.66 T to 1.15 T, respectively. The decrease in the out-of-plane saturation field is somewhat unexpected as it should be dominated by the thin-film shape anisotropy. However, a granular microstructure can lower the overall effective shape anisotropy making the film easier to saturate with an external magnetic field. Another potential explanation is the stabilization of HCP-Co could also lower the out-of-plane saturation field but this could not be confirmed X-ray diffraction due to the thin and disordered nature of the Co layers.

## (b) Intermixed Interface

To explore the effects of chemical intermixing at the interface we sputter deposited the multilayer stack Co(10 nm)/$Co_xPt_{1-x}$(1 or 2 nm)/Pt(3 nm) onto glass substrates. Note that for the study of the intermixing effect, on THz emission, all the layers including the spacers were deposited at 3 mTorr to form dense and smooth layers. The $Co_xPt_{1-x}$ layers were formed by Co-deposition and the values of x = 0.25, 0.50 and 0.75 were controlled by adjusting the deposition conditions.

## (c) Epitaxial FCC-Co and HCP-Co

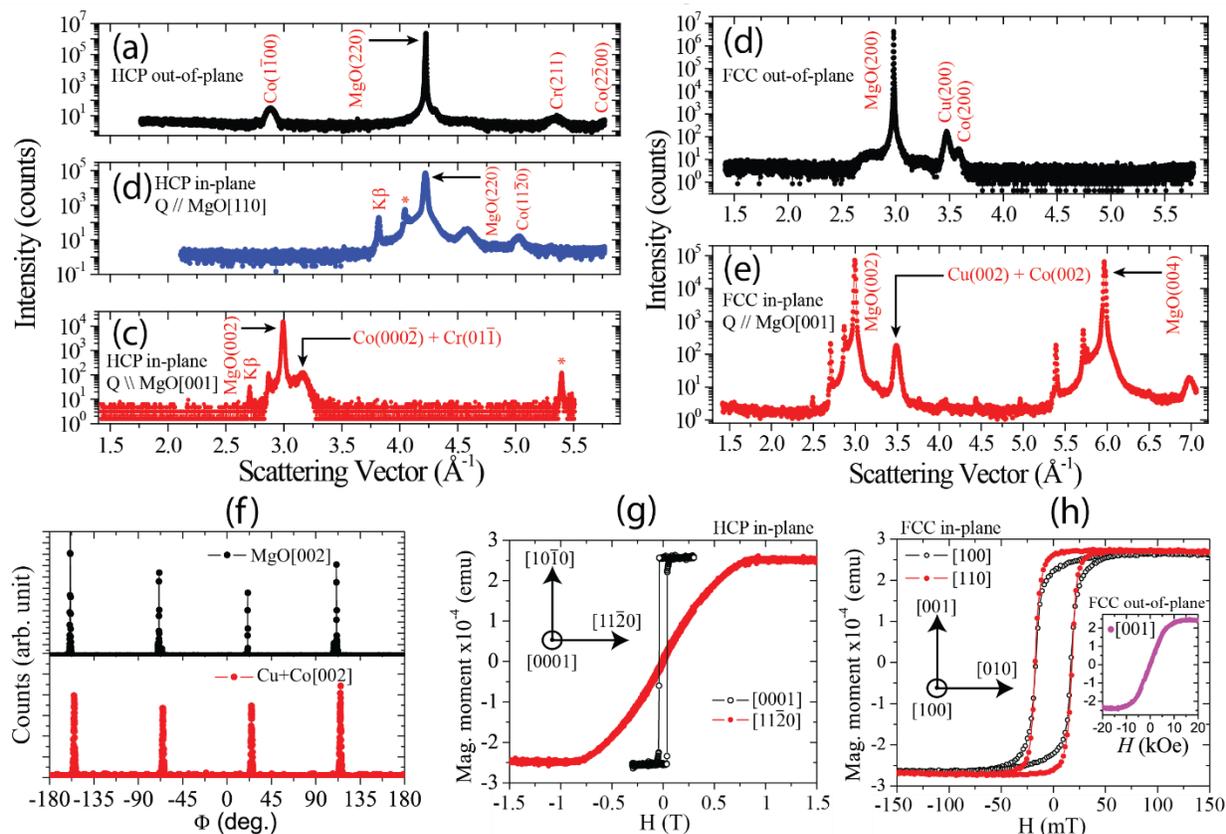

*Figure 3 The intensities of (a-e) are given in logarithmic scales. (a) The out-of-plane X-ray diffraction spectrum of the HCP-Co-film. (b) The in-plane spectrum where the scattering vector **Q** (the vector pointing along the bisection of the incoming and scattered beam) is along the MgO[110] direction, and (c) where **Q** is along the MgO[001] direction. The X-ray diffraction spectra of (d) the out-of-plane and (e) in-plane FCC-Co-film with **Q** along the MgO[001] direction. Note that the incident x-ray angle was fixed at $0.6^0$ from the surface when the in-plane spectra was recorded. (f) φ-scans of the MgO(002) (top) and Cu(002)+Co(002) (bottom) families of peaks. (g) In-plane Magnetic hysteresis loops of (g) the HCP-Co(10nm) and (h) FCC-Co(15nm)-film along various crystallographic directions.*

To explore the role of the crystal symmetry at the Co/Pt interface we grew epitaxial Co-films onto single-crystalline MgO substrates. Two particular lattice structures of Co were studied: HCP-Co(10-10) and FCC-Co(100)-films. The HCP-Co-films were grown on Cr(211)-buffered MgO(110) single-crystal substrates with the following structure: MgO(110)/Cr(5 nm)/Co(10 nm)/Pt(3 nm) [29]. The Cr-seed layer was grown at 300 °C into the body-centered cubic (BCC) structure oriented along BCC(211). The subsequent Co layer was deposited at 200 °C and capped with a 3 nm Pt-layer at room temperature, resulting in the easy axis of the magnetic anisotropy oriented in-plane along HCP-Co(0001). The epitaxial growth is confirmed by x-ray diffraction, the measurements are presented in Figs. 3(a-c) for both the out-of-plane and in-plane scattering geometries. The magnetic hysteresis loops for the in-plane and out-of-plane magnetic fields are shown in Fig. 3(g).

Using Cu(200)-buffered MgO(100) single-crystal substrates we obtained epitaxial FCC-Co with the following multilayer stacks MgO(100)/Cu(20 nm)/Co(10, or 15 nm)/Pt(3 nm). The Cu-seed layer was grown at 350 °C and oriented along FCC(200). The subsequent Co and Pt layers were grown at room temperature resulting in FCC-Co(200) oriented along the *a*-axis. The out-of-plane scattering of the 15 nm Co-film is shown with the Cu(200) and FCC-Co(200) peaks. The X-ray diffraction signals of both the FCC and HCP crystal structures are in good correspondence with those of the previously measured FCC-Co and HCP-Co in Ref. [30]. The magnetic hysteresis loops for the out-of-plane and in-plane magnetic fields are shown in Fig. 3(h). Note that prior to the deposition of the films the substrates were heated at 600 $^0$C for an hour under vacuum to obtain clean surfaces. Moreover, the base pressure throughout the deposition of all the samples was at $3 \times 10^{-8}$ Torr.

The differences in the optical absorption between the MgO and glass substrates for both the THz radiation and 800 nm laser pulse are small. Thus, the influence of different substrates on the performed THz experiments is negligible.

## (3) Experimental method

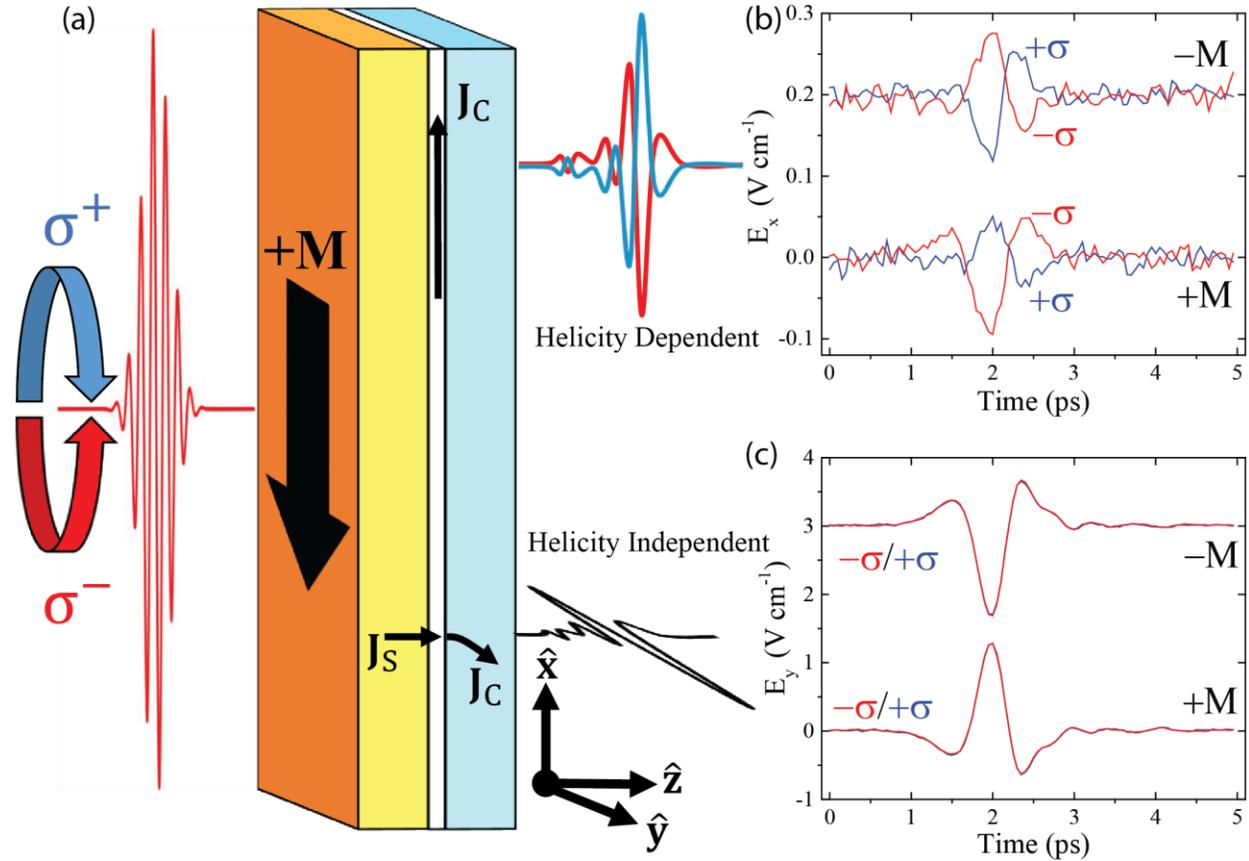

*Figure 4 (a) A schematic illustration for the laser induced helicity dependent and independent photocurrent generation. The graphs on the right show the time traces of the electric field of (b) the helicity-dependent (HD) and (c) helicity-independent (HI) THz emission exited by circularly polarized (±σ) light. In both cases, the THz electric field can be controlled by an external magnetic field of 120 mT which is enough to saturate the magnetization of Co. The figure shows how the reversal of the magnetization aligned along the $\hat{x}$-axis (±M) affects the time traces of the emitted THz electric field. The curves measured for –M are shown with an offset of 0.2 V/cm for $E_x$ in panel (b) and of 3 V/cm for $E_y$ in panel (c).*

A schematic depiction of the experimental geometry is shown in Fig. 4(a). To excite the photocurrents in the Co/Pt bilayer, we used a 40 fs circularly polarized laser pulse with a central wavelength of 800 nm and a pulse energy of 20 µJ, focused onto a 2-mm spot. An external magnetic field was applied in the plane of the sample to saturate its magnetization along the ±$\hat{x}$-axis (see coordinate system in Fig. 4(a)). The emitted THz radiation from the sample was focused onto a 1-mm-thick ZnTe crystal using gold coated parabolic mirrors. The time-resolved electric field was then obtained via electro-optical sampling in the 1-mm-thick ZnTe crystal. The setup is similar to the one described in Ref. [31] and all experiments were performed at room temperature.

We chose the coordinate system such that the propagation of the THz radiation was along the $\hat{z}$-axis and the external magnetic field was applied in-plane of the sample along the $\hat{x}$-axis. To measure the polarization of the THz radiation we employed two wire-grid polarizers. The axis of the first polarizer was fixed parallel to the $\hat{y}$-axis while the axis of the second polarizer was rotated ±45 degrees with respect to the axis of the first polarizer. This allowed us to determine the $\hat{x}$ and $\hat{y}$ components of the THz polarization. If the magnetization of the sample is aligned along the $\hat{x}$-axis, the THz emission from the HI and the HD photocurrents could be distinguished by detecting the THz signals (see Figs. 4(b,c)) with the electric field along the $\hat{y}$- and $\hat{x}$-axis, respectively.

The mechanism for generating the HI THz electric field is based on the generation of spin-polarized hot electrons in Co due to the heat driven ultrafast demagnetization [32]. This spin-polarized current flows from Co to Pt where it is converted into a charge current via the inverse spin-Hall effect [33,34]. The direction of the charge current **J**$_C$ is determined by the magnetization **M**, the spin-polarized current flow **J**$_s$, and the inverse spin-Hall angle α: $\mathbf{J_C} \propto \alpha \, \mathbf{J_S} \times \mathbf{M}$. The mechanism for the generation of the HD THz electric field is fundamentally different, since it requires circularly polarized light. The circularly polarized light will have a HD effect on the magnetization of Co, resulting in a spin-orbit torque acting on the spins of Co at the interface with Pt [35,36]. The breaking of inversion symmetry is essential for the coupling of the optically induced spin dynamics and the generation of interfacial HD photocurrents [37,38]. The direction of the photocurrent in this mechanism is determined by the magnetization **M**, the helicity of the light **σ**, and the direction along which the space inversion symmetry is broken (here given by the normal vector **n**): $\mathbf{J_C} \propto \mathbf{n} \times (\mathbf{M} \times \boldsymbol{\sigma})$.

In our analysis of the experimental data we retrieve the averaged HD THz ($E^{HD}$) and HI THz ($E^{HI}$) signals as

$$E^{HD} = \frac{1}{4}(E^{+\sigma+M} - E^{-\sigma+M} - E^{+\sigma-M} + E^{-\sigma-M}), \tag{1}$$

$$E^{HI} = \frac{1}{4}(E^{+\sigma+M} + E^{-\sigma+M} - E^{+\sigma-M} - E^{-\sigma-M}). \tag{2}$$

Here $E^{HD}$ and $E^{HI}$ are the polarized HD and HI THz electric field.

## (4) THz emission results

### (a) Interfacial Roughness Series

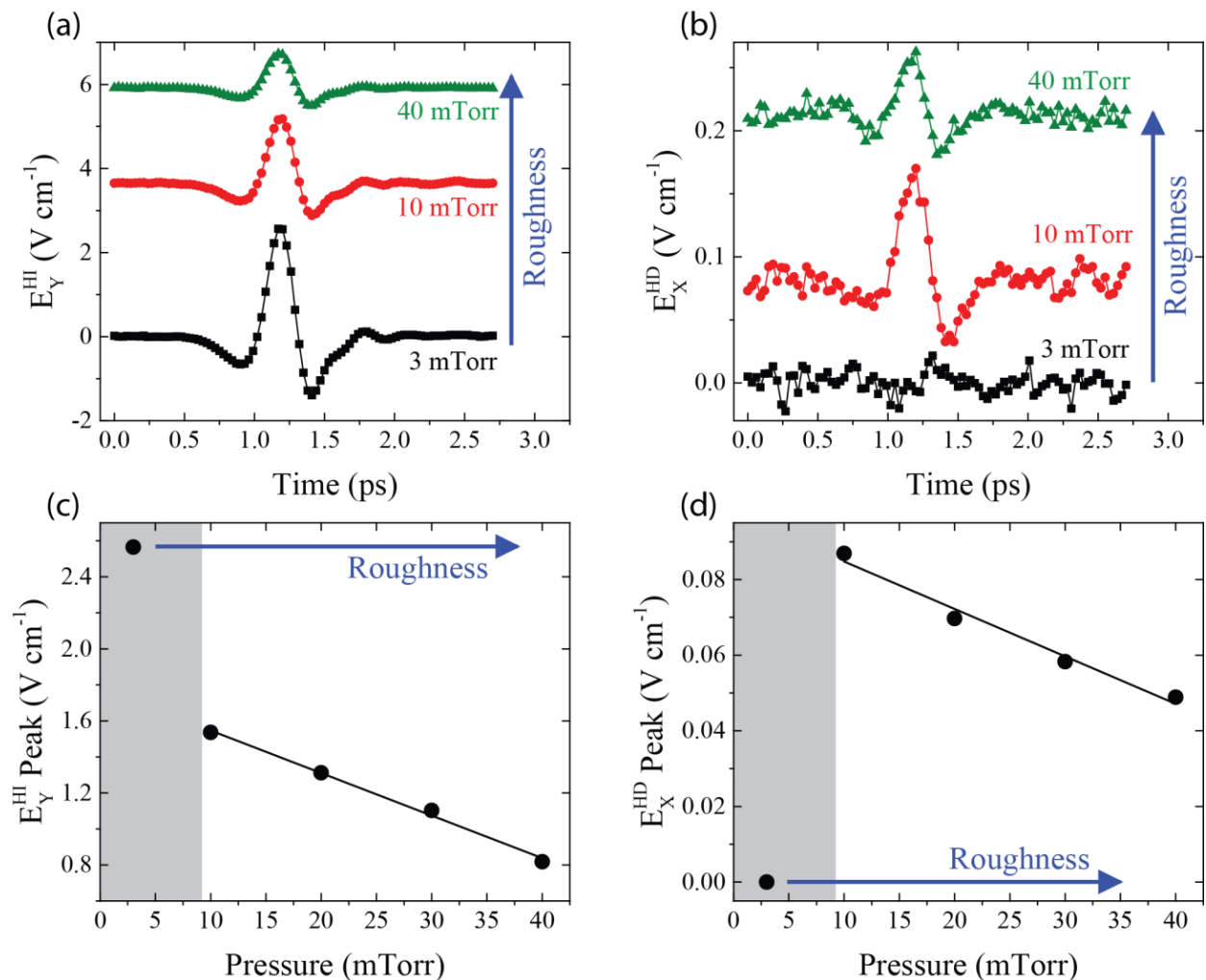

*Figure 5 (a) The helicity-independent and (b) helicity-dependent THz emission for various interfacial roughness values. The time traces are shown with a vertical off-set. The observed electric field at 0 ps is 0 for the measurements. The arrows indicate the increase of the interface roughness. Peak intensities of (c) the helicity-independent THz and (d) helicity-dependent THz emission. The gray area emphasizes deviations in the behavior. Beyond the gray area (>10 mTorr) both the helicity-independent and -dependent THz emission decreases linearly with increasing deposition pressures. The solid line is a guide to the eye.*

To probe the role of the sample structure on the generation of HD and HI photocurrents, we measured the THz electric field from the Co/Pt samples described above. We initially studied smooth polycrystalline Co/Pt bilayers grown at 3 mTorr on glass substrates (Fig. 2(a)). The results of the experiment are shown in Fig. 5 for HI THz (panel a) and HD THz (panel b), respectively. Similar to previous works, the figure reveals a strong HI signal [24,39]. However, the bilayer grown at 3 mTorr does not show any measurable HD THz signal (see Fig. 5(b)). An increase of the sputter pressure, that results in an increase of the structural disorder/interfacial roughness, changes the THz emission substantially. While an increase of the sputter pressure results in a decrease of the HI THz signal, an increase of the pressure from 3 mTorr to 10 mTorr results in a dramatic increase of the HD THz signal. Further increase of the pressure above 10 mTorr results in a decrease of the HD signal, similar to the HI signal. The peak amplitudes of the HI and HD THz signals are plotted against Ar deposition pressure in Figs. 5(c, d). The jump in HD THz signal upon an increase in pressure from 3 mTorr to 10 mTorr highlights the importance of roughness for the mechanism of HD THz emission.

## (b) Intermixed Interface

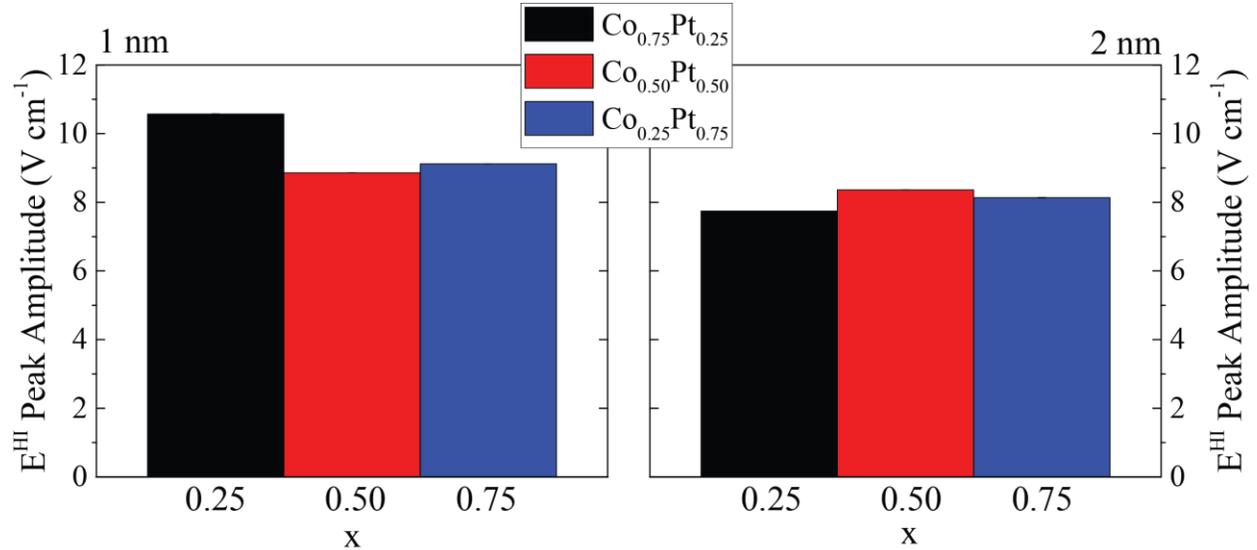

Figure 6 The peak amplitude of the helicity-independent THz emission from samples with the $Co_xPt_{1-x}$ alloy spacers with thicknesses (left panel) 1 nm and (right panel) 2 nm. The alloy compositions are $Co_{0.75}Pt_{0.25}$ (black), $Co_{0.50}Pt_{0.50}$ (red), and $Co_{0.25}Pt_{0.75}$ (blue). The mean absolute deviation which we chose as our error bar is negligible and therefore not visible in the graph.

To study the contribution of interfacial mixing we introduced a spacer layer of $Co_xPt_{1-x}$ alloy of various $Co_xPt_{1-x}$ compositions at the interface of the Co/Pt bilayers. The films were grown at 3 mTorr to have smooth interfaces, but with an alloy interlayer to simulate a diffuse interface. We employed alloys of $Co_{0.75}Pt_{0.25}$, $Co_{0.50}Pt_{0.50}$, and $Co_{0.75}Pt_{0.25}$ with thicknesses of 1 and 2 nm. The experimental procedure was similar to the one performed on the rough Co/Pt interfacial bilayers. Surprisingly, no HD THz emission was observed for all the samples having a $Co_xPt_{1-x}$ spacer layer. The HI THz emission was nevertheless strong for all six samples (see Fig. 6) and significantly higher than samples without an alloy interface. Note that during the sample growth only the concentration of Co and Pt and the sputtering time were varied, all other parameters were kept constant. The mean absolute deviation is taken by repeating the measurement three times, the extracted deviation value was less than 0.5% of the mean value and is therefore not visible in Fig. 6. The HI THz peak amplitude shown in Fig. 6 decreases with increasing the Co-Pt spacer layer thickness from 1 nm to 2 nm for all alloy composition. However, the HI THz emission does not show any correlations amongst the various Co-Pt compositions and spacer thicknesses. We compared the HI THz signal from samples with a $Co_xPt_{1-x}$ spacer and without a $Co_xPt_{1-x}$ spacer (Co/Pt bilayers with the roughest and the smoothest interfaces as well as with the epitaxial HCP-Co and FCC-Co layers), these results are shown in Table 1. We find a dramatic increase in the HI THz emission with insertion of an alloy interlayer. The largest increase was for the $Co_{0.25}Pt_{0.75}$(1nm) spacer which was 4.2 times stronger than the Co/Pt bilayer grown at 3 mTorr and 3.4 times that of the FCC Co/Pt bilayer. Even the weakest HI THz emission ($Co_{0.75}Pt_{0.25}$(2nm)) showed an amplification factor of 2.5 compared to the Co/Pt bilayer with the strongest HI THz emission (FCC Co/Pt).

## (c) Epitaxial FCC-Co and HCP-Co(10 nm)

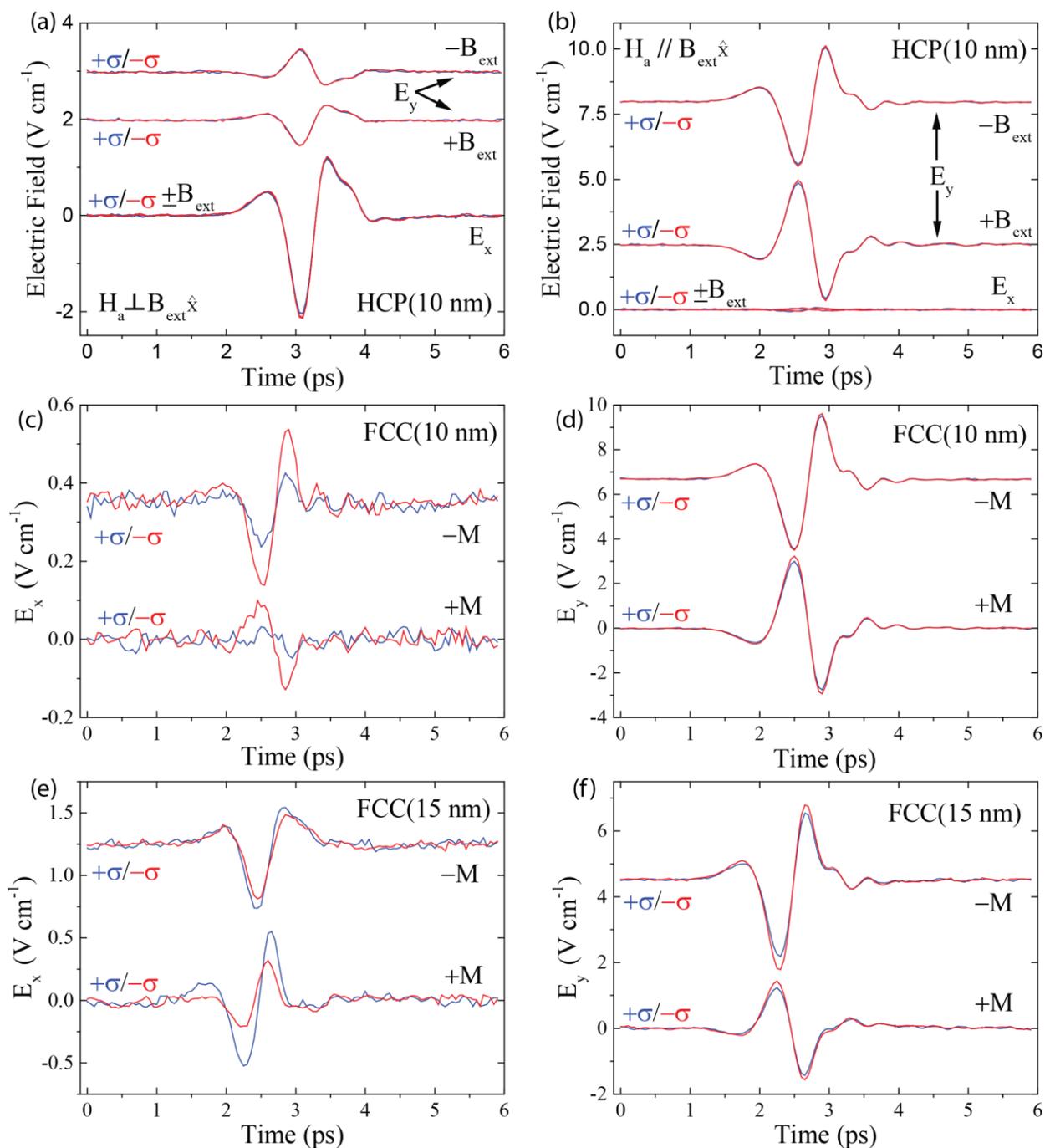

*Figure 7 Time traces of the THz electric field polarized along the $\hat{x}$ and $\hat{y}$ axes. The emission was generated by exciting Co/Pt bilayers with monocrystalline (HCP and FCC) Co with circularly (+σ/-σ) polarized laser pulses. The external magnetic field (±$B_{ext}$) is applied along the $\hat{x}$-axis at 100 mT. The easy axis of the in-plane magnetic anisotropy ($H_a$) of HCP-Co is aligned either (a) perpendicular or (b) parallel to the external magnetic field. The magnetic anisotropy field strength is 0.8 T. The THz emission polarized along the $\hat{x}$- and $\hat{y}$-axis for (c,d) FCC-Co(10 nm) and (e,f) FCC-Co(15 nm), respectively. For FCC-Co the magnetization is saturated and is aligned parallel along the external magnetic field.*

Finally, we studied the role of the crystal structure of Co layer in the process of THz emission. No HD THz emission was observed from the samples with monocrystalline Co, while the HI THz emission manifestation of the crystal structure of the Co-film was clearly seen. The crystalline orientation in our HCP-Co sample is (10-10). Due to the crystal structure, the material acquires an in-plane magnetic anisotropy with the hard and in-plane easy axis along the [11-20] and [0001] directions, respectively. We therefore studied the THz emission for two cases where the magnetic field was applied along [11-20] or [0001] directions. The magnetic anisotropy of Co had a remarkable effect on the THz emission. Figures 7(a, b) show the time traces of the THz electric field for the case when the external magnetic field ($B_{ext}$) is perpendicular and when it is parallel to [0001], the easy axis of the magnetic anisotropy ($H_a$). Note that the external magnetic field of 0.1 T was strong enough to saturate the magnetization along the easy axis. However, it was too weak to fully saturate the magnetization along the hard axis. The characteristic field of magnetic anisotropy was 0.8 T, estimated from the magnetic hysteresis loops shown in Fig. 3(g). The component of the THz electric field along the $\hat{y}$-axis ($E_y$), i.e. perpendicular to the external magnetic field ($B_{ext}$), reverses sign upon switching the polarity of the external magnetic field. At the same time, the component of the THz electric field along the $\hat{x}$-axis ($E_x$), i.e. parallel to external magnetic field, is insensitive to the polarity of the magnetic field. It shows that there is no contribution with the properties of the HD THz emission and thus no traces of photocurrent due to spin-dependent photogalvanic effects can be seen.

In the case of the FCC-Co, the magnetic film has a weak four-fold in-plane magnetic anisotropy with a strong out-of-plane shape anisotropy, i.e. with the hard-axis along the [001] direction (see Fig. 3(h)). The external magnetic field was also strong enough to saturate the in-plane magnetization. For this case we studied two FCC-Co layers with a thickness of 10 nm and 15 nm. Figures 7(c, d) show the time traces of the detected THz signal of FCC-Co(10 nm) with its polarization parallel ($E_x$) and perpendicular ($E_y$) to the external magnetic field, respectively. Similarly, Figs. 7(e, f) show the time traces of the THz signals of FCC-Co(15 nm). The THz signal polarized perpendicular to the external magnetic field ($E_y$) showed no clear difference between the FCC-Co(10 nm) and HCP-Co(10nm) samples. Moreover, although some sensitivity to the helicity of the femtosecond laser pulse is seen, the polarity of the THz electric field does not change upon reversing the light helicity. Therefore, also in this case we conclude that spin-dependent photogalvanic effects play a minor role if any.

Table 1 *The peak amplitude of the helicity-independent THz emission from Co/Pt heterostructures with the roughest (40 mTorr) and smoothest (3 mTorr) interfaces, Co/Pt with $Co_xPt_{1-x}$ alloy interfaces, and epitaxial FCC-Co and HCP-Co interfaces.*

| Sample | HI Peak Amplitude [V/cm] |
|---|---|
| Co(10nm) / Pt(3nm)    (40 mTorr) | 0.8 |
| Co(10nm) / Pt(3nm)    (3 mTorr) | 2.5 |
| Co(10nm) / Pt(3nm)    (HCP) | 2.44 |
| Co(10nm) / Pt(3nm)    (FCC) | 3.15 |
| Co(10nm) / $Co_{0.75}Pt_{0.25}$(2nm) / Pt(3nm) | 7.74 |
| Co(10nm) / $Co_{0.50}Pt_{0.50}$(2nm) / Pt(3nm) | 8.36 |
| Co(10nm) / $Co_{0.25}Pt_{0.75}$(2nm) / Pt(3nm) | 8.13 |
| Co(10nm) / $Co_{0.75}Pt_{0.25}$(1nm) / Pt(3nm) | 9.12 |
| Co(10nm) / $Co_{0.50}Pt_{0.50}$(1nm) / Pt(3nm) | 8.86 |
| Co(10nm) / $Co_{0.25}Pt_{0.75}$(1nm) / Pt(3nm) | 10.57 |

## (5) Discussion

### (a) Interface roughness

By varying the deposition pressure during the Co growth process, we have demonstrated that one could control the microstructure and interfacial roughness of Co/Pt bilayers. From the static magnetometry data, shown in Fig. 2, we observed that the change of the growth pressure also affects the magnetic properties of the Co-film. An increase in the coercivity of the in-plane magnetization together with a decrease in the out-of-plane saturation magnetization was observed.

Interestingly, interface roughness has opposite effects on the HI and HD THz emission from Co/Pt bilayer (Figs. 5(a, c)). Upon changing the pressure from 3 to 10 mTorr, the strength of the electric field of the HI THz emission drops by factor of 2, while the electric field of the HD counterpart experiences an increase from full suppression. It is clear that both the HI and the HD THz emission must depend on the optical properties of the bilayers at the wavelength of the femtosecond laser pulse (800 nm), transport properties, and the magnetization of the Co-film in a similar way. However, such a dramatic difference in the behavior of the HI and the HD THz emission allows us to exclude changes of optical, transport, and magnetic properties of the Co-film as reasons for the difference in the THz emission from bilayers grown at 3 mTorr and 10 mTorr. Moreover, if any changes of the growth procedure would affect the spin-orbit interaction in Co, one would expect a change of the characteristic times of the laser-induced demagnetization of the Co-film [42]. The latter must lead to a change of the temporal profile of the THz emission. Our experiments reveal that the temporal profiles of the electric field for both HI and HD THz emission are similar for samples with different roughness. Therefore, it allows us to exclude a change of laser-induced magnetization dynamics as a reason for the difference in the THz emission from bilayers grown at 3 mTorr and 10 mTorr. All these findings indicate that the totally different trends of the HI and the HD THz emissions upon the pressure increase must be related to opposite effects of the interface roughness on the efficiency of photocurrent generation via inverse spin-Hall and spin-dependent photogalvanic effects, respectively.

It is expected that roughness and or other types of disorder at the Co/Pt interface must lead to an increase of the spin-flip probability. Therefore, the interface becomes less transparent for spin currents from Co to Pt layers and the strength of the charge photocurrent generated in the Pt-layer due to the inverse spin-Hall effect will decrease. Such a decrease can explain the observed weakening of the HI THz emission upon an increase of interface roughness caused by the increase of the growth pressure from 3 to 10 mTorr.

In order to explain the dramatic increase of the HD THz emission upon the transition from a smooth to a rough interface, we note that the mechanism of THz emission can be seen as a nonlinear optical phenomenon similar to Second Harmonic Generation (SHG). The latter can also be efficiently generated from interfaces of metallic bilayers. For instance, it was shown that the SHG intensity increases with an increase of the roughness of a CoNi/Pt interface [43]. Similar to these studies, we explain here the increase of the efficiency of the HD THz emission as a result of geometrical increase of the volume with the properties of the interface. An effective thicker interface layer, where Co is in direct contact with Pt, results in a larger net photocurrent and thus stronger THz emission.

## (b) Intermixing

The experimental results show that the HI THz signal generated in Co/Pt bilayers with a $Co_xPt_{1-x}$ spacer is much stronger compared to the HI THz signal generated in Co/Pt with various roughnesses or crystal structures (see Fig. 6 and Table 1). The weakest HI THz emission signal of the trilayer (Co/Co$_{75}$Pt$_{25}$(2nm)/Pt) was 2.5 times stronger than the highest HI THz emission signal of a bilayer (FCC Co/Pt). Interestingly, intermixing has a very different effect on the HD THz emission. After an introduction of an intermixed alloy layer, the efficiency of the HD THz emission dropped down to the noise level. Again, although intermixing can affect many bilayer properties, which are important for laser-induced THz emission, such a difference in the trends of the HI and the HD counterparts indicate that the intermixing most likely affects the generation of photocurrents.

We propose two possible explanations for the enhancement of the photocurrent due to the spin-Hall effect. Firstly, the introduction of a $Co_xPt_{1-x}$ alloy layer could generate a higher influx of spin currents injected into Pt due to a reduction of spin dissipation and decoherence at the interface [40, 41]. The separation of the Co and Pt layer by a $Co_xPt_{1-x}$ alloy spacer can possibly reduce the proximity effect that Co has on Pt as well as reducing the interfacial spin resistivity due the sharp transition between Co and Pt. Secondly, it is possible that the $Co_xPt_{1-x}$ spacer layers have a significant spin-orbit interaction, which increases the spin to charge current conversion. This interface-dependent mechanism for the optimization of the THz emission has not been discussed before and thus reveals new opportunities for further increase of the intensity of the THz radiation from Co/Pt and similar multilayers [10-14].

The observed reduction of the HD THz emission emphasizes the importance of a direct contact between pure Co and Pt for the generation of photocurrents due to the spin-dependent photogalvanic effect.

## (c) Crystal structure

Finally, although our experiments with epitaxially-grown Co-films revealed that also the crystal structure of Co may influence the HI THz emission, the effect is relatively small. Moreover, no HD THz emission was observed from the samples with monocrystalline Co. Therefore, from these experiments we conclude that crystal structure of the Co-film has no large effect on the spin-current through the Co/Pt interface nor on the charge current in the Pt layer due to the inverse spin-Hall effect. Regarding the fact that polycrystalline Co more easily forms rough interfaces with Pt, the results also emphasize the importance of rough Co/Pt interfaces for photocurrent generated due to the spin-dependent photogalvanic effect.

## (6) Conclusion

In conclusion, using THz emission from Co/Pt bilayers grown under various conditions we studied how photocurrents generated in the bilayer depend on the properties of the Co/Pt interface. Inducing a relatively high interfacial roughness at the interface of Co/Pt is crucial for the generation of helicity-dependent THz emission and its associated photocurrent at the interface due to the spin-dependent photogalvanic effect. Simultaneously, the roughness results in a decrease of the transparency of the interface for the spin-current from the Co to Pt layer. As a result, photocurrents in the Pt layer due to the inverse spin-Hall effect and polarization-independent THz emission will also decrease. The study of the THz emission from Co/Pt samples with $Co_xPt_{1-x}$ spacer layers at various $Co_xPt_{1-x}$ compositions and thicknesses showed that the intermixing of Co/Pt at the interface does not help to improve the efficiency of the spin-dependent photogalvanic effect and the intensity of the helicity-dependent THz emission. However, the intermixing appears to play a significant role in amplifying the helicity-independent THz emission generated due to the inverse spin-Hall effect. The observed HI THz emission from the $Co/Co_xPt_{1-x}/Pt$ trilayers were enhanced by a factor of 4.2 for $Co_{25}Pt_{75}$(1nm) compared to the HI THz emission from the Co/Pt bilayer grown at 3mTorr. Finally, we observed a dramatic suppression of the photocurrent due to the spin-dependent photogalvanic effect which resulted in a full suppression of the helicity-dependent THz emission in Co/Pt with crystalline FCC or HCP-Co. Therefore, these observations show that the photocurrent due to the spin-dependent photogalvanic effect is the largest for the cases of rough Co/Pt interfaces with polycrystalline Co and no intermixing of Co and Pt, while the spin current from Co to Pt is larger for smooth interfaces. The experiments also show that either the spin current or its conversion to the charge current together with the associated helicity-independent THz emission can be enhanced by intermixing between Co and Pt layers. In none of the THz emission mechanisms, the crystal structure of Co seems to play a significant role.

## (7) Acknowledgements

This work was funded by DOE grant # DE-SC0018237, the Nederlandse Organisatie voor Wetenschappelijk Onderzoek (NWO), the European Union Horizon 2020 and innovation program under the FET-Open grand agreement no.713481 (SPICE), the European Research Council ERC grant agreement no.339813 (EXCHANGE), the ministry of Education and Science of the Russian Federation project no.14.Z50.31.0034, and the Russian Science Foundation 16-12-10520. We want to thank Ray Descoteaux from CMRR, Sergey Semin, Tonnie Toonen, and Chris Berkhout from Radboud university for all the technical support.

## (8) References